# Recent Development in Analytical Model for Graphene Field Effect Transistors for RF Circuit Applications

Abhishek Kumar Upadhyay[1*], Ajay K. Kushwaha[2], Deepika Gupta[3], Santosh K. Vishvakarma[4]

[1]Institute for Fundamentals of Electrical Engineering and Electronics, Technische Universität Dresden, Germany, 01069
[2]Discipline of Metallurgy Engineering and Materials Science, Indian Institute of Technology Indore, Simrol, Indore, M.P., 453552
[3]Department of Electronic and Communication Engineering, IIIT Naya Raipur, Atal Nagar Chhattisgarh, 493661
[4]Discipline of Electrical Engineering, Indian Institute of Technology Indore, Simrol, Indore, M.P., 453552

*Corresponding: meetabhishek14@gmail.com;

**ABSTRACT**

The MOS devices are the basic building block of any digital and analog circuits, where silicon (Si) is the most commonly used material. The International Technology Roadmap Semiconductor (ITRS) report predicts the gate length of the MOS device will shrink to 4.5 nm up to 2023, this may create severe short channel effects (SCEs). Therefore, new channel materials have been realized, which have shown their potential to maintain the proper balance between device performance and SECs. Among them, graphene has shown its strong presence as an alternative channel material in terms of its fascinating electrical and mechanical properties. It has ultra-high carrier mobility (77,000 cm²V⁻¹s⁻¹) and saturation velocity (4.5 ×10cm s⁻¹), which makes it compatible with high-speed circuit applications. This review paper has a detailed report on the several analytical modeling approaches for graphene-based FET device, which includes the drift-diffusion, gradual channel approximation, virtual source, and ballistic approaches. The device modeling plays an important role to predict the device performance and used to reveal the physics behind it. In addition, the compact model of the device will use in the development of electronic design automation (EDA) tools, which are used for the circuits simulation.

## 1. INTRODUCTION

In order to achieve high-speed analog/radio frequency (RF) circuits, aggressive downscaling of semiconductor MOS devices is required. In addition, the scaling of MOS device gives further benefits of low-power consumption, low power-delay product, high density as well as high processor speed. However, the downscaling of these devices may leads to several short channel effects (SCEs) such as drain-induced barrier lowering (DIBL), surface scattering, velocity saturation and hot carrier effects etc. Importantly, to mitigate these kinds of SCEs, different scientific and academic communities have investigated various novel device structures and channel materials. These MOS devices with novel structure and channel material has introduced as the non-conventional MOS devices. These non-conventional MOS devices have high level of physics hence; the device modeling for such devices is complex and has moved from classical to quantum level.

Furthermore, this paper is going to discuss a nonconventional MOS device with graphene channel material. It is a two dimensional (2D) atomic layer of carbon atoms with a honeycomb lattice structure. Recently, graphene has attracted the enormous attention as a channel material for MOS devices due to its extraordinary electrical and mechanical properties [1]-[8]. In addition, graphene as channel material helps the MOS downscaling need as the conventional channel material scaling has reached to its ultimate limit hence imposes several restrictions on semiconductor device scaling.

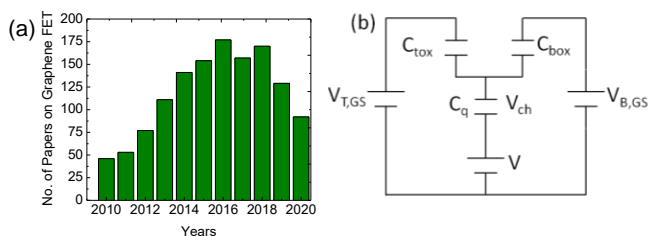

Figure 1. (a). Total number of papers on Graphene FET based experimental and theoretical demonstrations electrical characteristics published per year and listed on Web of Science [9]. (b) Equivalent capacitive circuit for the modeled GFET

Interestingly, many works have been published depicting the modeling of graphene material, MOS device characteristics with graphene as channel material and its use in VLSI circuits.

The compact model of MOS device plays a vital role for accurate implementation of VLSI circuit from the semiconductor MOS devices. Kacprzak et al. reported the first DC compact model of GaAs-FETs in 1983 [10], after which the compact modeling for FETs started receiving much attention to explore physics of the MOS device. Interestingly, compact models of MOS devices are essential for the development of electronic design automation (EDA) tool hence; circuit simulation. In addition, the electrical characteristics of the MOS device can also be alternatively investigated by the compact modeling approach [11]. Therefore, compact modeling of MOS device also helps to analyze the performance and stability of any integrated circuits (IC).

Moreover, for graphene FET many modeling approaches such as drift-diffusion (DD), gradual channel approximation (GCA), and virtual source (VS) modeling approach are commonly used to formulate the electrical parameters reported in [12]-[34]. The mathematical expression proposed by these compact models are simplified enough for the implementation of graphene FET in existing circuit design environments using hardware description languages (HDL) like Verilog-A. Importantly, authors in [21]-[23] presented an analytical model that define the boundary points between the regions to ensure Jacobian continuity.

The large area graphene channel based FET, makes it compatible for analog/RF applications with the cutoff frequencies ($f_T$) in the Gigahertz range. Note that some RF circuits have been fabricated and demonstrated with GFET, such as a RF mixer, frequency multiplier and amplifier circuits [35]-[38].

## 2. ELECTRICAL& MECHINAL PROPERTIES

The Graphene is a single atomic thick sheet of sp$^2$ hybridized carbon atoms having unique material properties in regards to many electrical and mechanical applications. Importantly, it has superior electrical-thermal conductivity, optical transparency and chemical stability over other semiconducting materials. Also, It exhibits many other desirable properties including, high thermal conductivity (5000 Wm$^{-1}$K$^{-1}$), and the large critical current densities ($\sim$3 × 10$^9$ A/cm$^2$) [1]-[6]. Further, the atomic thickness of graphene monolayer sheet is ideally utilized in the large-area nano manufacturing of wearable and flexible electronic devices. In addition to electrical superiority, it also possesses excellent mechanical strength in terms of the strain limit of material. Apart from this, graphene also exhibits the highest carrier mobility of any known material, which is 100 times greater than the Si as well as better than other high mobility 2D semiconductors. Owing to the smaller effective mass of the charge carriers, it can be seen that graphene affords the highest charge carriers mobility atrelatively high carrier density (e.g.$\sim$10$^{12}$/cm$^2$).

However, this carrier mobility in any semiconductor FET device only describes the charge transport speed in low electrical fields. Whereas, the state of the art scaled FET devices shows very high channel electrical field than conventional device. This causes steady-state carrier velocity to saturate in the device. At this point, the intrinsic carrier mobility becomes less relevant for device performance, and the carrier saturation velocity becomes more important factor in the charge transport characteristics. In this regard, the graphene exhibits ultrahigh carrier saturation velocity. The maximum carrier velocity with graphene is expected to reach around 4.5 × 10$^7$ cm s$^{-1}$. In addition, under high fields, the carrier velocity in graphene does not drop as severely as those in III-V semiconductors. Therefore, the extraordinarily high carrier transport makes it an attractive material for ultra-high speed flexible RF electronics.[39]-[53].

### 2.1 E-k Relationship

E-k relationship describes the association between the energy and momentum of available quantum mechanical states for electrons in the semiconductor material. The E-k relationship for graphene in the first Brillouin zone (BZ) is given by [16]

$$E(\mathbf{k}) = E - E_{cv} = s\hbar vF|\mathbf{k}| \qquad (1)$$

where s= +1 for the conduction band (CB) and s=-1 for the valence band (VB), $\hbar$ is the reduced Plank's constant, $v_F$is the Fermi velocity of the charge carrier in Graphene, $|\mathbf{k}| = \sqrt{k_x^2 + k_y^2}$ is the wave vector of the charge carrier in two dimensional (2D) monolayer graphene sheet, At $|\mathbf{k}| = 0$ is the point where minima of CB is meet with maxima of VB called Dirac Point. Due to the zero band gap the conduction band minima($E_C$) and the valence band maxima($E_V$) coincide, i.e., $E_C = E_V = E_{CV}$.

### 2.2 Density of State

The density of state (DoS) in any semiconductoris definedas the number of states per interval of energy at each energy level available to be occupied [13],[77].Hence, the number of available states $N$ in $k$ space for graphene, considering the valley and spin degeneracy factor, is given by [14]

$$N = g_s g_d \frac{\pi k^2}{(2\pi/L)^2} = g_s g_d \frac{Ak^2}{4\pi} \qquad (2)$$

or,

$$N = g_s g_d \frac{A}{4\pi} \left( \frac{E - E_{cv}}{\hbar v_f} \right)^2 \qquad (3)$$

where $g_s$ and $g_d$ are the spin and valley degeneracy factors respectively, with $g_s = g_d = 2$, and $L$ the length of the graphene sheets. Now using Equation (3), the DOS per unit area $D(E)$ for graphene is obtained as,

$$D(E) = \frac{1}{A}\frac{dN}{dE} \qquad (4)$$

Putting Equation (3) into Equation (4), we have the DOS

$$D(E) = \frac{2}{\pi} \frac{|E - E_{cv}|}{\left(\hbar v_f\right)^2} \qquad (5)$$

The density of state is basically used to derive the expressions for the voltage dependent hole and electron sheet carrier densities in the channel region of graphene FET.

### 2.3 Carrier concentration

The voltage-dependent carrier density of the hole in the graphene channel is given by [13],

$$p = \int_{-\infty}^{E_{cv}} D(E)[1 - f(E)]dE \qquad (6)$$

where, $f(E)$ is the Fermi Dirac distribution, now putting DOS from Equation (5) to Equation (6) the hole carrier sheet density is given by,

$$p = \int_{-\infty}^{E_{cv}} \frac{2}{\pi} \frac{|E - E_{cv}|}{\left(\hbar v_f\right)^2} [1 - f(E)]dE \qquad (7)$$

Now changing the order of integration of above equation can be written as

$$p = \int_{-E_{cv}}^{\infty} \frac{2}{\pi} \frac{|E + E_{cv}|}{\left(\hbar v_f\right)^2} [1 - f(E)]dE \qquad (8)$$

Using $E_{cv}$ as a reference energy i.e.for $E_{cv} = 0$, $E_f = q \times V_{ch}$ is the Fermi energy. Here, $V_{ch}$ is the channel potential drop across the graphene channel. It can be formulated using Fig. 1(b) [14],

$$V_{ch}(x) = \left(V_{T,eff} - (V_{DS}/2)\right) \frac{C_{tox}}{C_{tox} + C_{box} + 0.5C_q} \quad (28)$$
$$+ \left(V_{B,eff}\right.$$
$$\left. - (V_{DS}/2)\right) \frac{C_{box}}{C_{tox} + C_{box} + 0.5C_q}$$

where, $C_{tox}$, $C_{box}$ is the capacitance per unit area formed by top, back oxide. $C_{TOX} = \frac{\epsilon_{tox,box}}{t_{ox}}$, where $t_{tox,box}$ is the thickness of top, back gate oxide layer and $\epsilon_{tox,box}$ is permittivity of the top and back oxide material. $V_{T,eff} = \left(V_{TGS} - V_{TGS,0}\right)$ and $V_{B,eff} = \left(V_{BGS} - V_{TGS,0}\right)$, where are the top and back gate over drive voltages respectively. $C_q$ is the quantum capacitance, defined as the derivative of the net channel sheet charge density $(Q_{net})$ with respect to $V_{ch}$ is given as [54]-[60],

$$C_q = -\frac{dQ_{net}}{dV_{ch}} \quad (25)$$

The minus sign stems from the fact that a more positive gate voltage may leads to a more negative charge in the graphene channel.

The net carrier density $(Q_{net})$ in the graphene sheet, participates in the conduction of GFET is equal to the difference in between hole sheet density and electron sheet density and is given as [13],

$$Q_{net} = q \times (p - n) \quad (11)$$

$$Q_{net} = \frac{2q}{\pi(\hbar v_f)^2} \left[ \begin{array}{c} \int_0^\infty \frac{E}{\exp\left(\frac{E + E_{cv}}{k_B T}\right) + 1} dE \\ - \int_0^\infty \frac{E}{\exp\left(\frac{E - E_{cv}}{K_B T}\right) + 1} dE \end{array} \right] \quad (12)$$

The above equation gives the net charge carrier density participated in the current conduction of GFET but the carrier density due to thermal excitation is still underestimated close to the *Dirac* point [13]. Following the fact that in zero-bandgap graphene material holes, electrons and residual charge carrier density additively contribute to the overall current, the new equation for the sheet density in graphene FET is given as [18],

$$Q_{tot} = q \times (p + n) \quad (13)$$

or,

$$Q_{tot} = \frac{q\pi(K_B T)^2}{3(\hbar v_f)^2} + \frac{q^3 V_{ch}^2}{\pi(\hbar v_f)^2} + q n_{pud} \quad (14)$$

where, $n_{pud}$ is residual charge for monolayer graphene sheet, which is due to electron-hole puddles is given by [61]

$$n_{pud} \approx \frac{2}{\pi \hbar^2 v_f^2} \left( \frac{\Delta^2}{2} + \frac{\pi^2}{6} k_B^2 T^2 \right) \quad (15)$$

It is observed that the total charge density improve the accuracy of the GFET models because of residual charge density for small $(qV_{ch} \ll KT)$ and larger value $(qV_{ch} \gg KT)$ of $V_{ch}$ is taken into accounts [15]. The channel carrier density is used to calculate the quantum and different parasitic capacitances formed at different interface of Graphene FET by using Mayer and charge conservation approach. These approaches are described in detail through the next section.

## 2.4 Capacitances
In case of Graphene FET basically, two type of approach is used to the calculate capacitances formed at different interface of Graphene FET: 1) Mayer technique 2) Charge conservation techniques

### 2.4.1 Mayer and Mayer Like Approach:
This approach is widely used because of its simplicity and fast computation. However, in this technique, it is assumed that the capacitances in the intrinsic FET are reciprocal. Considering this, the channel charge is given by [13]-[16], [18]

$$Q_{CH}(x) = W \int_0^L \left(Q_{net/tot}(x) + q n_{pud}\right) dx \quad (16)$$

The above equation is simplified by the changing the integration variable and limits of the integration we have new equation is given by,

$$Q_{CH}(V) = \frac{qW}{E_{av}} \int_0^{V_{DS}} \left(Q_{net/tot}(x) + q n_{pud}\right) dV \quad (17)$$

Where $E_{av} = \frac{dV}{dx} \approx \frac{V_{DS}}{L}$ is the electric field along the channel for long channel GFET.

#### 2.4.1.1 Gate-Source Capacitance $C_{gs}$
The small-signal gate-source capacitance can calculated as,

$$C_{gs} = \frac{dQ_{CH}}{dV_{gs}}\bigg|_{V_{ds}=const.} \quad (18)$$

#### 2.4.1.2 Gate-Drain Capacitance $C_{ds}$
The small-signal gate-source capacitance can calculated as,

$$C_{ds} = \frac{dQ_{CH}}{dV_{ds}}\bigg|_{V_{gs}=const.} \quad (19)$$

Although the Meyer model exhibit well-known problems that the capacitances in the intrinsic FETs are reciprocal, which is not valid in case of real devices, and this model cannot ensure charge conservation [37].

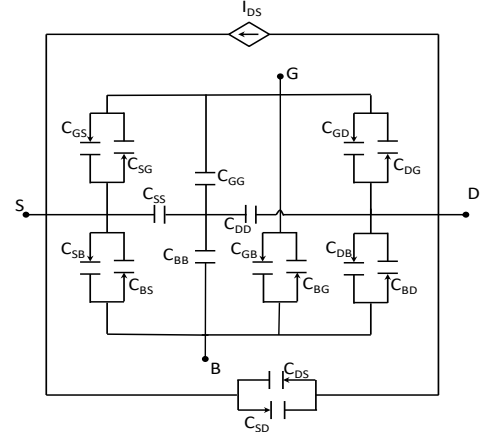

Figure.3 Ward-Dutton's equivalent capacitive model of GFET device.

### 2.4.2 Charge conservation techniques
The charge-based technique ensures the charge conservation and considers the nonreciprocity of the capacitances in MOS devices. These features are required especially for RF circuit applications in which the influence of transcapacitances is important and must be taken into account. The charge based capacitance modeling for four-terminal FET can have fourself capacitances and twelve intrinsic transcapacitances, which makes 16 capacitances in total as shown in Fig. (3) [62].

The capacitance matrix is formed by these capacitances where each element $C_{ij}$ describes the dependence of the charge at particular terminal $i$ with respect to a varying voltage applied to terminal $j$, assuming that the voltage at any other terminal remains constant [63]-[65].

$$C_{ij} = -\frac{\partial Q_i}{\partial V_j} \text{ for } i \neq j \text{ and } C_{ij} = \frac{\partial Q_i}{\partial V_j} \text{ for } i \neq j$$

Where, $i$ and $j$ stands for S, D, G and B

$$\begin{bmatrix} C_{GG} & -C_{GD} & -C_{GS} & -C_{GB} \\ -C_{DG} & C_{DD} & -C_{DS} & -C_{DB} \\ -C_{SG} & -C_{SD} & C_{SS} & -C_{SB} \\ -C_{BG} & -C_{BD} & -C_{BS} & C_{BB} \end{bmatrix} \quad (21)$$

The summation of the elements of the each row and column must sum to zero to preserve the charge conservation principle of the given device. It is also notice that the in 16 intrinsic capacitances only 9 are independent remaining are the dependent capacitances.

## 3. MODELING OF GFETs

The fundamental goal of a device modeling is to obtain the functional relationship among the electrical terminals of the MOS devices that is to be modeled. The electrical characteristics depend upon the set of parameters including device dimensions and the physics involved in in device. In the following section, we are going to discuss few of the approaches to model GFET successively:

### 3.1 Drift-Diffusion(DD) Approach:

In this section, we are going to discuss the modeling of drain current ($I_{DS}$) for monolayer GFET based on drift-diffusion (DD) approach. In general, the $I_{DS}$ of a MOS devices is given by [12]-[19]

$$I_{DS} = -q\rho_{sh}(x)v(x)W \quad (29)$$

where, $\rho_{sh}(x)$ is the free carrier sheet density in the channel at position $x$, $v(x)$ is the carrier drift velocity, $V(x)$ is the potential along the channel length ($L$).

Monte Carlo simulations have shown that the steady-state velocity can be approximated by the expression [18],

$$v = \frac{\mu E}{1 + \frac{\mu|E|}{v_{sat}}} \quad (30)$$

where $E$ is the electric field, $\mu$ is the low-field mobility, and $v_{sat}$ is the saturation velocity. If graphene channel is located on $SiO_2$, the $v_{sat}$ is strongly affected by the two interfacial surface optical phonons of $SiO_2$ with energies of $59 meV$ and $155\ meV$. We follow this approach by introducing a correction term $A \times V^2(x)$, the saturation velocity is [14]

$$V_{sat} = \frac{\Omega}{(\hbar \rho_{sh})^{0.5} + AV^2(x)} \quad (31)$$

Combining (29), and (30), the $I_{DS}$ becomes

$$I_{DS} = \frac{q\rho_{sh}\mu(-dV/dx)}{1 + \frac{\mu(-dV/dx)}{v_{sat}}} \quad (32)$$

where, $V = V(x)$. Here, (32) solved by the separation of variables, integrating the left-hand-side over $x$ from $x = 0$ to $x = L$, and RHS over $V(x)$ from $V(0) = 0$ to $V(L) = V_{DS}$, the final expression of the $I_{DS}$ [14],[15]

$$I_{DS} = q\mu W \frac{\int_0^{V_{DS}} \rho_{sh} dV}{L - \mu \int_0^{V_{DS}} \frac{1}{v_{sat}} dV} \quad (33)$$

This model is used to explain the detailed device physics; however, not suitable for implementation in analog circuit modeling languages, such as SPICE or Verilog-A. So, Rodriguez et.al., [16].There are distinct mobility for electron and hole in graphene [19], [66]-[68]. In addition, it has also observed that carrier mobility decreases with the increase in the carrier density in the monolayer graphene. So the effective carrier mobility($\mu_{eff}$) is [19],

$$\mu_{eff} = \frac{n\mu_n + p\mu_p + n_{pud}h}{n + p + n_{pud}} \quad (34)$$

Where $\mu_n$ and $\mu_p$ is the mobility of the electron and hole respectively, $h = (\mu_n + \mu_p)/2$

C. Mukherjee et. al. have explored bias stress to discuss the failure mechanisms that are to be originated from the generation of traps and interface states causing a shift in the transfer characteristics and mobility degradation, respectively in monolayer graphene [69]. For the development of the aging compact model, the trap density is implemented in the pre-stress compact model to modulate the channel potential is given by [20].

$$N_{Trap} = N_{SS}\left[1 - exp\left(-\left(\frac{t}{\tau}\right)^\delta\right)\right] \quad (37)$$

where $N_{SS}$ parameter needs to be a function of the stress voltage. Here, $N_{SS}$ can be defined as $N_{SS} = N_1 + N_2 V_{GS,stress}^{\alpha 1} - N_3 V_{BG,stress}^{\alpha 2}$ where, $N_1$, $N_2$ and $N_3$, $\alpha 1$ and $\alpha 2$ are the fitting parameters. The DD approach is well mature to capture the all regions of device operation of GFET and requires the assumption of carrier density-dependent saturation velocity to reproduce the experimental characteristics. Further, in order to avoid this complexity during device modeling, B. W. Scott et. al. suggested the gradual channel approximation approach to calculate the drain current for GFET [21].

### 3.2 Gradual Channel Approximation Approach

The gradual channel approximation (GCA) and generally is valid for long-channel MOSFETs, with the assumption that charges carrier ($Q(x)$) in the surface depletion region are induced solely by the gate voltage. Here, the $Q(x)$ is the electric charge density along the channel from source to drain and can be given by

$$Q(x) = -C_{top}[V_{g0} - V(x)] \quad (37)$$

The transport characteristics of the GFET are modeled by splitting the carrier distribution function into it's even and odd parts  i.e.$f(k) = f_{even}(k) + f_{odd}(k)$. Now, under randomizing collisions, and in high fields, the Boltzmann transport equation is given by [21]-[23]

$$\frac{eF}{\hbar} \frac{\partial}{\partial k_x} f_{even}(\vec{k}) = -\frac{1}{\tau_{tot}(k)} f_{odd}(\vec{K}) \quad (38)$$

With$1/\tau_{tot}(k) = \Sigma_i 1/\tau_i(k)$. Here, $i$ indicate different scattering mechanism and $F$ is the electric field. Further, the thermalized even carrier distribution in presence of strong inter-carrier scattering for high carrier concentration is,

$$f_{even}(\vec{k}) = \frac{1}{1 + exp[(\hbar v_f)(k - k_f)/(K_B T)]} \quad (39)$$

Where, $k_f = kF(x)$ is carrier concentration along the channel. Importantly, for p-type GFET, the current can be calculated as

$$\vec{I} = \frac{4e}{L} \sum_{\vec{k}} \vec{v}(\vec{k}) f_{odd}(\vec{k}) \quad (40)$$

where the factor 4 accounts for the spin and the two fold degeneracy of the $Dirac$ point [4], $v(\vec{k}) = v_F(cos\theta, sin\theta)$ and $\theta$ is the angle between the $F$ and the vector k. For $\hbar v_F k_F \gg k_B T_E$ one can approximate $\left(\frac{\partial}{\partial k_x}\right) f_{even}(\vec{k})$ by a delta function centered around $k_F$. After integration, and for given $k_F = \sqrt{\pi p}$ [67], the $I_{DS}$ can be calculated as

$$I_{DS} = W \frac{2e^2}{h} F v_F \tau(p) \sqrt{\pi p} \quad (41)$$

Where, $p$ is the hole concentration, and $\tau(p)$ is the relaxation time. Specifically, we assume $\tau(p) = \tau_{lf}/(1 + F/F_C)$ where $F_C$ is the critical field for the high field regime [67], $\tau_{lf}(p) = \tau_0\sqrt{p/N_i}$ is the low-field relaxation time dominated by scattering with charged impurities with density Ni [67], and $\tau_0$ is a time constant. By setting $\mu_0 = (e/\hbar) v_f \tau_0 \sqrt{p/N_i}$, one recovers the conventional current expression as

$$I_{DS} = Wepv(F) \quad (42)$$

With, $v(F) = \mu_0/(1 + F/F_C)$. Further, for the charge-control model, close to the $Dirac$ point, one can use the mass action law [70] to get

$$p(x) = \frac{Q(x)}{2e} + \sqrt{\left(\frac{Q(x)}{2e}\right)^2 + p_0^2} \quad (43)$$

Where, $p_0$ is the minimum sheet concentration $V_{g0} = V_{gtop} - V_0$ where $V_0$ is the threshold voltage is define as [15]

$$V_0 = V_{gtop}^0 + \frac{C_{back}}{C_{top}}\left(V_{gback}^0 - V_{gback}\right) \quad (44)$$

Where, $V_{gtop}^0$ and $V_{gback}^0$ designate the top and back gate voltages at the $Dirac$ point, respectively. For $\frac{Q(x)}{2e} \gg p_0$, which the case for all bias conditions is considered in this analysis, one can write

$$p(x) = \frac{Q(x)}{e} \quad (45)$$

Now, integrating the (42) from source to drain as in conventional MOS devices, and by taking into account the series resistance $R_s$ at the source and drain terminal,

$$I_{DS} = \frac{W\mu_0 V_C}{2LC_{top}(|V_{DS}| - 2|I_{DS}|R_s + V_C)}[Q(L)^2 - Q[0]^2] \quad (46)$$

Where, $Q(L) = -C_{top}(V_{g0} - V_{DS} - |I_{DS}|R_S)$ and $Q(0) = -C_{top}(V_{g0} - |I_{DS}|R_S)$. In addition, the low drain-source bias conductance is readily calculated by taking the derivative of $I_{DS}$ with respect to $V_{DS}$ as $V_{DS} \to 0$.

$$g_{ds}(V_{ds} \to 0) = \frac{-V_{g0}}{|2R_S V_{g0} - R_C V_C|} \quad (47)$$

Where, $1/Rc = (W/L) \mu_0 C_{top} V_c$, so that $R_c V_c$ is independent of $V_c$, as is the conductance at low drain bias. The low drain source bias resistance,

$$R_{ds} = \frac{1}{g_{ds}} = 2R_S - \frac{R_C V_C}{V_{gs0}} \quad (48)$$

It establishes a linear relation between $1/g_{ds}$ and $\frac{1}{V_{gs0}}$ with a slope given by $R_c V_c$ (inversely proportional to the mobility) and an asymptotic conductance value for large $V_{g0}$ reaching $2R_s$. In the same context, one the $V_{DS(sat)}$ as a function of the top gate voltage $V_{g0}$ by solving for $V_{ds}$ after setting the derivative of the $I_{DS}$ with respect to $V_{ds}$ equal to zero that yields

$$V_{DS(sat)} = \frac{2\gamma V_{g0}}{(1 + \gamma)^2} + \frac{1 - \gamma}{1 + \gamma^2}\left[V_c - \sqrt{V_c^2 - 2(1 + \gamma)V_c V_{g0}}\right] \quad (49)$$

With $\gamma = R_s/R_c$. Substituting the (50) into the current equation (47) enables us to obtain the expression of the saturation drain ($I_{DS(sat)}$)

$$I_{DS(sat)} = \frac{\gamma}{R_s(1 + \gamma)^2}\left[-V_c + (1 + \gamma)V_{g0}\right. \\ \left. + \sqrt{V_c^2 - 2(1 + \gamma)V_c V_{g0}}\right] \quad (50)$$

By taking the derivative of the saturation current with respect to the top gate voltage, one derives the expression for the transconductance at saturation

$$g_m^{sat} = \frac{1}{R_s + R_c}\left[1 - \frac{1}{\sqrt{1 - 2(1 + \gamma)V_{g0}/V_c}}\right] \quad (51)$$

Additionally, the expression for the electric potential as a function of position along the channel length can be derived from the current Equation (47) and is given by

$$V(x) = V_{g0} - V_i + \sqrt{\left(V_{g0} - V_i - I_{ds}R_s\right)^2 + \frac{2I_{ds}x}{W\mu_0\mu_0 C_{tox}}} \quad (52)$$

where, $V_i = I_{DS}/W\mu_0 F_C C_{top}$ and the source is located at $x = 0$. This paper extends the virtual-source (VS) model to provide intrinsic-charge (capacitances) descriptions that, for the first time, extend all the way to the ballistic regime, where it can be shown that the GCA is often violated.

### 3.3 Virtual Source Model

In Virtual Source model [27]-[30], the FET current is given as the product of areal charge density $Q_{x0}$ at the virtual source and carrier injection velocity ($v_{x0}$). In single layer graphene, has two virtual sources one for electrons and another for holes at opposite ends of the channel because of its gapless nature.

The net drain current ($I_{DS}$) in the ambipolar virtual source model is a superposition of the injected electron and hole currents, is given by

$$\frac{I_{DS}}{W} = (Q_{x0e} + Q_{x0h})v_{x0}F_{sat} \quad (53)$$

where, $Q_{x0e}$ and $Q_{x0h}$ are the electron and hole concentrations, respectively, at the virtual source points can be determined numerically using the $Fermi\text{-}Dirac$ integral and DOS.

$$Q_{x0e} = \int_0^\infty DoS(E) \frac{1}{1 + exp\left(\frac{E - qV_{CS}}{k_BT}\right)} dE \quad (54)$$

and

$$Q_{x0h} = \int_0^\infty DoS(E) \frac{1}{1 + exp\left(\frac{E + qV_{CD}}{k_BT}\right)} dE \quad (55)$$

where,

$$V_{CS} = \frac{(V'_{GS} - V_{min,0})C_G}{C_G + C_q(V_{CS})} \quad (56)$$

and

$$V_{CD} = \frac{(V'_{GS} - V_{min,0})C_G}{C_G + C_q(V_{CD})} \quad (57)$$

where, $v_{x0}$ is assumed to be identical for electrons and holes due to the band symmetry in graphene. Here, $F_{sat}$ is an empirical function used for the transition in current from the linear to the saturation region, given as

$$F_{Sat} = \frac{V'_{DS}/V_{dsat}}{\left(\left(1 + \left(\frac{V'_{DS}}{V_{dsat}}\right)\right)^\beta\right)^{1/\beta}} \quad (58)$$

$$V_{dsat} = \frac{v_{x0}L}{\mu} \quad (59)$$

where,$V'_{DS}$ is the intrinsic drain-source bias, $\mu$ is the carrier mobility (assumed identical for electrons and holes), $L$ is the channel length, and $\beta$ is an empirical parameter obtained upon calibration with experimental data.

### 3.4 McKelvey's Flux Theory

The McKelvey's flux theory (MFT) and quasi ballistic transport model (BTM) both are used to obtain an compact analytical drain current expression for GFETs. The beginning of the channel is defined at the location of virtual cathode, where the electric potential reaches to its minimum or the barrier height for electrons reaches to the maximum [71]. Specifically, the flux is defined as the carriers that pass through the channel per unit width per second. Also, the $I_{DS}$ is given by charges carried by net fluxes between the positive and negative going fluxes as follows [72]-[73],

$$I_{DS} = Wq[F^+(0) - F^-(0)] \quad (60)$$

where, $F^+(0)$ and $F^-(0)$ are positive and negative going carrier fluxes, respectively. Moreover, the source-injected fluxes, $F^+(0)$ can be expressed as

$$F^+(0) = \bar{v}_x (K_B T)^2 \Im_1(\eta_F) / \pi (\hbar v_f)^2 \quad (61)$$

where $\Im_1(\eta_F)$ is the Fermi dirac integral, $\eta_F = (E - E_F)/k_BT$, $E_F = qV_{CH}$, $V_{CH}$ is the voltage drop across the quantum capacitance, $\hbar$ is the reduced Plank's constant, $v_f$ is the Fermi velocity, $K_B$ is the Boltzmann constant. Similarly, $F^-(0)$ is the negative going flux generated from drain end due to thermionic emission is,

$$F^-(0) = rF^+(0) + (1-r)F_b^-(L) \quad (62)$$

Due scattering, the negative going flux ($F^-(0)$) at drain end divided into two parts. The first one is due to backscattering of charge carriers in the channel $rF^+(0)$ rt is due to the carrier injection from drain side, i.e. $(1-r)F_b^-(L)$, $F_b^-(L)$ is the negative directed flux is caused by thermal emission from the drain and is given by

$F_b^-(L) = F^+(0)e^{-qV_{DS}/K_BT}$, where symbols have already defined.

Therefore, the drain current is calculated by using Equation (62) and (61) in Equation (60), we have

$$I_{DS} = \frac{qWV_Tn(x)(1-r)[1 - \Im_1(\eta_F - U_{DS})/\Im_1(\eta_F)]}{1 + r + (1-r)\,\Im_1(\eta_F - U_{DS})/\Im_1(\eta_F)} \quad (63)$$

where, $n(x)$ represents the charge carrier density of the channel is given by [34].

$$n(x) = \left[\beta \left[\frac{q^2V_{CH}}{\pi(\hbar v_f)^2}\left(V_{CH} - \frac{V_{DS}}{2}\right)\right] \right.$$
$$\left. + (1-\beta)\left[\frac{t_{gr}\epsilon_{gr}}{q}\left(\xi - \frac{\phi}{\lambda^2}\right)\right] + N_{imp}\right] \quad (64)$$

where $\lambda = \sqrt{\frac{\lambda_T^2 \lambda_B^2}{\lambda_T^2 + \lambda_B^2}}$ is the scale length, $\lambda_T^2 = \frac{\epsilon_{gr}t_{ox}t_{gr}}{\epsilon_{ox}}$, $\lambda_B^2 = \frac{\epsilon_{gr}t_{box}t_{gr}}{\epsilon_{box}}$, $t_{gr}$ is the thickness of graphene, $\epsilon_{gr}$ is the dielectric permittivity of graphene, $\xi = ((V_{T,eff} - V_{T,fb} + V_{B,eff} - V_{B,fb})/\lambda^2)$, $V_{T,fb}/V_{B,fb}$ is the flat band voltage of top/bottom gate, and it is assumed to be zero, $V_{T,eff} = V_{T,GS} - V_{T,GS0}$, $V_{B,eff} = V_{B,GS} - V_{B,GS0}$, $V_{T,GS}/V_{B,GS}$ is the top/bottom gate voltage, and $V_{T,GS0}$ and $V_{B,GS0}$ are the Dirac voltages, which are used as fitting parameters

The thermal velocity in case of pure ballistic transport is called injection velocity i.e. $V_{inj} = V_{Th} = \sqrt{(2K_BT)/\pi m^*}$, $m^*$ is effective mass of the charge carrier. However, in case of quasi ballistic transport where the back scattering take place among the carrier, the thermal (injection) velocity is given by the $V_{inj} = V_{Th,2D} \cdot \bar{T}$, where $\bar{T}$ is being the net transmission coefficient and $V_{Th,2D}$ is the thermal velocity of the charge carrier is given as [59]

$$V_{Th,d} = V_{Th}\frac{\Gamma[(d+1)/2]}{\Gamma(d/2)} \quad (65)$$

Where,$V_{Th}$ is the thermal velocity, $d$ is the dimension of the material. For 2D material the thermal velocity ($V_{Th,2D}$) is given by

$$V_{Th,2D} = V_{Th}\frac{\sqrt{\pi}}{2} \quad (66)$$

The averaged transmission coefficient ($\bar{T}$) between source and drain terminal of GFET is [71]

$$\bar{T} = e^{\frac{1}{2}(\bar{\alpha})x}[Coshqx + (\bar{\alpha}/q)Sinhqx]^{-1} \quad (67)$$

where, $x = L$ is the channel length of the GFET, $\bar{\alpha} = \lambda^{-1} + q|E|/K_BT$ is the scattering probabilities of the charge carriers.

Interestingly, the back scattering coefficient ($R$), is the most important parameter for the quasi ballistic transport model and heavily biased towards the source side. Note that with previously published papers the same back scattering coefficient [32], were utilized to describe source-to-drain and drain-to-source backscattering. However, the backscattering coefficients are different for source to drain and drain to source, and are dependent on electric field of the channel region. The model for the back scattering has developed by considering the field independent mean free path, as well as the differing scattering mechanism are also not taken into account. So in order to overcome these issue a new expression for the backscattering coefficient is proposed in [33], is by

$$R = \frac{2\left[\frac{D}{qE_x}ln\left(1+\frac{L}{L_{KT}}\right)+C\langle E_{2D}\rangle L\left(1+\frac{L}{L_{KT}}\right)\right]}{\pi v_f + 2\left[\frac{D}{qE_x}ln\left(1+\frac{L}{L_{KT}}\right)+C\langle E_{2D}\rangle L\left(1+\frac{L}{L_{KT}}\right)\right]}$$ (68)

For long channel (under low field) device, $L/L_{KT} \to 0$.

$$R = \frac{2\left[\frac{D}{\langle E_{2D}\rangle}+C\langle E_{2D}\rangle\right]L}{v_f\pi + 2\left[\frac{D}{\langle E_{2D}\rangle}+C\langle E_{2D}\rangle\right]L}$$ (69)

Where, $L_{KT} = \langle E_{2D}\rangle/qE_x$ is the distance, at which the potential drops to a value of $K_BT/q$, $C\langle E_{2D}\rangle$ is the in-plane phonon scattering rate and $D/\langle E_{2D}\rangle$ is the flexural phonon scattering rate.

The quasi-ballistic mobility was introduced in [72], [73] with minor difference. Here, authors in [72] assume that the ballistic velocity is both appropriate for 3D and unidirectional, but only half Maxwellian is directed towards the channel. On the other hand, authors in [73] use it as a full Maxwellian velocity for non-degenerate regime which is not valid for degenerate regime and is valid for only 3D materials.

However, in quasi-ballistic regime the conventional mobility equation is replaced by the effective mobility ($\mu_{eff}$), modeled by Mattestain's rule and given as follows [72]

$$\frac{1}{\mu_{eff}} = \frac{1}{\mu_0} + \frac{1}{\mu_B}$$ (70)

where, $\mu_b = \frac{qL}{m^*v_{Th,2D}}$ is the ballistic mobility of the charge carrier and $\mu_0$ is the physical or scattering limited mobility, while rest other symbols have their usual meanings.

**3.5 Landauer Approach**

This modeling approach is based on $\lambda$ obtained from fitting the experimental low-field $\sigma(n_s)$ characteristics with analytical expressions derived from the Landauer equation. The Landauer equation for the steady-state current is given by [74][75]

$$I_{DS} = \frac{1}{q} \int G(E)[F_1(E) - F_2(E)]dE$$ (71)

where, $G(E)$ is the energy-dependent conductance function, $F_1(E) = F(E, E_{F1})$ and $F_2(E) = F(E, E_{F2})$ denote the Fermi functions at the source and drain contacts that account for the difference in Fermi levels resulting from the applied bias as given by $qV_{DS} = (E_{F2} - E_{F1})$. Here, the conductance function is given by

$$G(E) = \frac{2q^2}{h} M(E)T(E)$$ (72)

Where, $M(E)$ and $T(E)$ are the energy-dependent density of modes, and the transmission coefficient respectively. The difference in the Fermi functions can be simplified for low bias conditions using the Taylor series expansion of $f(E, E_F)$ to obtain $(f_1 - f_2) \approx (-\partial f/\partial E)(qV_{DS})$. Using this simplification, and combining (71) and (72) allows expressing the low-field conductance as [31],

$$G = \frac{I_{DS}}{V_{DS}} = \frac{2q^2}{h} \int M(E)T(E)\left(-\frac{\partial f}{\partial E}\right)dE$$ (73)

Importantly, the density of modes is obtained analytically using the linear band structure of graphene, Figure 1(b) and is given by $M(E) = 4W|E|/(\hbar v_f)$. Here, W is the width of the graphene channel and $v_f$ is the constant electron velocity($\approx$ $10^8 cm/s$). Note that for low bias, we can express the transmission coefficient as $T(E) = \lambda(E)/[\lambda(E) + L]$, where $\lambda(E)$ is the energy-dependent backscattering mean free path. Substituting $M(E)$ and $T(E)$ into Equation (73) results in a general expression for low-field conductance in graphene given by

$$G = \frac{2q^2}{h} \int \frac{\lambda(E)}{\lambda(E) + L}\left(\frac{4WE}{hv_f}\right)\left(-\frac{\partial f}{\partial E}\right)dE$$ (74)

where, symbols have their usual meanings.

## 4. CONCLUSION

The current status of the analytical modeling of graphene-based FETs (GFETs) have been reviewed in this paper. However, it is also observed that, there are many challenges for digital applications of GFETs, while analog applications may have brighter future. So in order to simulate the different circuits and device structure the electronics automations tools are required. The device modeling has very much important role in the development of the EDA tools. In doing so, the techniques that are used by the different research groups to capture the different device region of operations. The drift-diffusion approach that is enough mature to capture all the characteristics of the single and double gate GFET device, having the assumption of carrier density-dependent saturation velocity. While the GCA is a charge-control model, fails to capture the ballistic nature of graphene. The first time, extend all the way to the ballistic regime in the virtual source approach. The detailed discussion of the ballistic nature of graphene FET is discussed using the Landauer Approach having its own limitation is terms of the coherency of the charge carrier and MFT approach. In addition, all the discussed modeling approach is simplified enough, so it can be implemented in the Verilog-A, which is suitable for a RF/analog circuit design. These circuits have been shown the benchmark against high-performance and ambipolar electronics circuits.